\documentclass{article} 
\usepackage{iclr2024_conference,times}

\usepackage{amsmath,amsfonts,bm}

\def\eqref#1{equation~\ref{#1}}

\def\1{\bm{1}}

\DeclareMathAlphabet{\mathsfit}{\encodingdefault}{\sfdefault}{m}{sl}
\SetMathAlphabet{\mathsfit}{bold}{\encodingdefault}{\sfdefault}{bx}{n}

\usepackage{hyperref}
\usepackage{url}

\usepackage[utf8]{inputenc} 
\usepackage[T1]{fontenc}    
\usepackage{hyperref}       
\usepackage{url}            
\usepackage{booktabs}       
\usepackage{amsfonts}       
\usepackage{nicefrac}       
\usepackage{microtype}      
\usepackage{xcolor}         
\usepackage{multicol}
\usepackage{multirow}
\usepackage{threeparttable}
\usepackage{adjustbox}
\usepackage{array} 
\usepackage{booktabs} 
\usepackage{amssymb} 
\usepackage{xspace}
\usepackage{enumitem}

\title{DRDT: Dynamic Reflection with Divergent Thinking for LLM-based Sequential Recommendation}
\author{%
  Yu Wang \\
  University of Illinois Chicago\\
  \texttt{ywang617@uic.edu} \\
  \And
  Zhiwei Liu\thanks{Corresponding Author: zhiweiliu@salesforce.com}\\
  Salesforce AI Research\\
  \texttt{zhiweiliu@salesforce.com} 
  \And
  Jianguo Zhang, Weiran Yao, Shelby Heinecke\\
  Salesforce AI Research \\
  \texttt{\{jianguozhang, weiran.yao, shelby.heinecke\}@salesforce.com}
  \\
  \AND
  Philip S. Yu  \\
  University of Illinois Chicago \\
  \texttt{psyu@uic.edu} \\
}

\newcommand{\modelname}{\textsf{DRDT}\xspace}
\iclrfinalcopy 
\begin{document}

\maketitle

\begin{abstract}

The rise of Large Language Models (LLMs) has sparked interest in their application to sequential recommendation tasks as they can provide supportive item information. However, due to the inherent complexities of sequential recommendation, such as \textit{sequential patterns across datasets}, \textit{noise within sequences}, and the \textit{temporal evolution of user preferences}, existing LLM reasoning strategies, such as in-context learning and chain-of-thought are not fully effective. To address these challenges, we introduce a novel reasoning principle: \textbf{Dynamic Reflection with Divergent Thinking within a retriever-reranker framework}.

Our approach starts with a collaborative in-context demonstration retriever, which collects sequences exhibiting collaborative behaviors as in-context examples. Following this, we abstract high-level user preferences across multiple aspects, providing a more nuanced understanding of user interests and circumventing the noise within the raw sequences. The cornerstone of our methodology is dynamic reflection, a process that emulates human learning through probing, critiquing, and reflecting, using user feedback to tailor the analysis more effectively to the target user in a temporal manner.

We evaluate our approach on three datasets using six pre-trained LLMs. The superior performance observed across these models demonstrates the efficacy of our reasoning strategy, notably achieved without the need to fine-tune the LLMs. With our principle, we managed to outperform GPT-Turbo-3.5 on three datasets using 7b models e.g., Vicuna-7b and Openchat-7b on NDCG@10. This research not only highlights the potential of LLMs in enhancing sequential recommendation systems but also underscores the importance of developing tailored reasoning strategies to fully harness their capabilities.
\end{abstract}

\section{Introduction}


The remarkable advancements in Large Language Models (LLMs) also inspire the exploration of LLM-based recommender systems.
The effectiveness of LLM-based recommender systems stems from vast pre-trained knowledge in LLMs, such as the understanding of item information and explanations of user behaviors~\cite{gao2023chat, liu2023chatgpt, hou2023largellmrank,wang2023zero}. 
One approach involves fine-tuning existing LLMs to tailor them to recommendation scenarios~\cite{bao2023tallrec, chen2023palr, lin2023rella, cui2022m6, wu2023exploring}, with methods like TALLRec~\cite{bao2023tallrec} using LORA~\cite{hu2021lora} technology for this adaptation with minimal training examples. 
Another strategy is fine-tuning LLMs specifically for recommendation tasks~\cite{geng2022recommendationp5, hou2023learning, hou2022towards, li2023textisallyouneed}.
Additionally, some research regards LLMs as zero-shot rankers~\cite{liu2023llmrec, gao2023chat, liu2023chatgpt, wang2023zero}, where they directly rank candidate items based on contextual information, \textit{e.g.} past user interactions. 
However, the strategies for prompting LLMs are not yet fully explored, particularly in more complex areas like sequential recommendation, which consider the evolving nature of user interactions. Specifically, current methods involve using historical interactions to formulate prompts about possible future engagements in the form of: \textit{The user has such historical interactions: <history>, what is the possible next engagement among the candidate set <candidates>?} In the realm of zero-shot LLM ranking, further investigations~\cite{hou2023largellmrank} have delved into in-context-learning (ICL) and chain-of-thought (COT)~\cite{wei2022chain} approaches for prompt construction. An example includes using historical interactions, ending with the penultimate item as a query, and the second-last item as an answer, coupled with a "step-by-step" thought process to facilitate LLM's recommendation process. These varied approaches highlight the ongoing exploration and potential of LLMs in enhancing the effectiveness of recommendation systems.

The significant impact of prompt strategy on the reasoning capabilities of Large Language Models (LLMs) is the primary motivation for this paper. Recognizing the under-explored area of reasoning principles in sequential recommendation, this study aims to introduce and evaluate various prompting strategies to fully harness the reasoning potential of LLMs for sequential recommendation tasks without fine-tuning process, thereby aiming to bridge existing gaps in the field. To establish an effective reasoning framework for sequential recommendation using LLMs, it is essential to address these three challenges:

\begin{enumerate}[leftmargin=*] 
\item The first challenge lies in \textit{capturing the collaborative signal for recommendations}. Conventional recommender systems make recommendations by referring other collaborative users with similar behaviors. In sequential recommendation, these signals are learned through similar sequence representations. However, existing prompting strategies, including ICL and COT, are predominantly informed by current sequences, lacking a view across the broader dataset that includes inter-item correlations. Moreover, the context length limitations of LLMs make it impractical to feed entire datasets into the LLM during inference.

\item The second challenge is \textit{handling the personalized aspects and ubiquitous noise}. Different users may care about different aspects when consuming the same item. Furthermore,  user interactions with items can often include a certain amount of noise, which does not accurately reflect their true preferences.

\item The third and perhaps most intricate challenge is accurately \textit{capturing the temporal evolution of user interests}. Unlike general NLP tasks like question-answering, sequential recommendation requires understanding how a user's preferences change over time, a nuance that current prompting strategies like plain-text prompting, ICL, and COT are not inherently designed to handle.

\end{enumerate}

Addressing the identified challenges, our paper introduces a novel reasoning principle called \textbf{D}ynamic \textbf{R}eflection with \textbf{D}ivergent \textbf{T}hinking (\modelname) within a retriever-reranker framework. This approach is designed to effectively incorporate collaborative signals, manage personalized aspects and noise, enable temporal reasoning and omit error accumulating caused by hallucinations.
\paragraph{Collaborative In-Context Demonstration Retriever:} 
The ICL within LLM-based recommender systems~\cite{hou2023largellmrank} primarily uses a portion of the target sequence to create an in-context example. However, this approach mainly aids the LLM in understanding the task and output format without offering additional supportive information. To enhance this process, we introduce a collaborative in-context demonstration retriever. This component is designed to gather sequences that end with the same item as the target user's collaborative in-context demonstrations. The in-context example informs LLM about the choice made by this collaborative user next. By employing this method, we not only assist the LLM in grasping the nature of the task but also provide it with reference information about the choices made by other users in similar situations. 

\paragraph{Divergent Thinking for Various Aspects and Noise Management:}  

Existing COT implementations predominantly follow a convergent thinking paradigm. In zero-shot COT, the phrase "Let's think step-by-step" is often added to prompt the LLM to autonomously develop a reasoning path. Typically, the LLM tends to use the similarity between a candidate and a historical item as one of the reasoning paths. In few-shot COT, specific examples of reasoning paths are provided. Once a path is established, results are generated along it. Variants like COT-SC and TOT construct multiple reasoning paths, determining results through a consensus among these paths. However, in the context of sequential recommender systems, constructing predefined reasoning paths before inference presents challenges. Users consider multiple aspects – such as price, color, reviews, quality, etc. – and different users may prioritize these aspects differently for the same target items. Additionally, noise within sequences is ubiquitous. If this noise is not accurately identified and filtered out, the recommender system might misinterpret it as evidence of diverse interests, rather than discerning clear patterns in a user's true preferences. To address these challenges, we propose a shift from the traditional convergent thinking paradigm to a divergent thinking paradigm. Instead of seeking a universal reasoning path or allowing the LLM to independently construct a path, our approach involves considering user engagement from multiple aspects. Recommendations are then made based on this multi-faceted analysis, rather than solely relying on similarity comparisons between candidate and historical items. This method aims to provide a more nuanced and accurate reflection of individual user preferences and interests.

\paragraph{Dynamic Reflection for Temporal Reasoning:} 

Existing prompting strategies often fall short of capturing the temporal evolutions inherent within a sequence. Specifically, ICL primarily assists LLMs in understanding the task's nature and in formulating the output. While COT encourages the LLM to reason through a certain path step-by-step, it often faces difficulty in identifying the most relevant factors leading to current user engagements. This uncertainty can lead to \textit{hallucination}~\cite{yao2023llmlies, martino2023knowledge, zhang2023siren}, where the LLM makes assumptions that deviate from the actual influencing factors, resulting in unreliable recommendations and potential error accumulation along the reasoning path. Instead of generating answers along an unreliable path, we propose a dynamic reflection scheme inspired by human learning processes, which includes probing, critiquing, and reflecting. 
Dynamic reflection begins with using a segment of the target sequence to prompt the LLM to probe the next possible item via divergent thinking. This process involves a two-step approach: firstly, the LLM generates a multi-aspect analysis to understand the various dimensions of user preferences. Following this, it utilizes this analysis to predict a specific item, based on the information available in the preceding subsequence.
We then use this item to prompt LLM to critique its prediction and the associated analysis. This step allows the LLM to adjust and correct its analysis, thereby preventing the accumulation of errors. Following this, the process moves to the next item in the temporal order, with repeated rounds of probing, critiquing, and reflecting. Through this iterative process, the LLM can dynamically uncover the true preferences of the users. 

The effectiveness of the \modelname principle was evaluated using three datasets and six existing LLMs. The numerical results demonstrate the effectiveness of this approach. We hope this paper can inspire further research in designing effective prompts for sequential recommendation systems. Our contributions can be summarized as follows:
 \begin{itemize}[leftmargin=*] 
    \item To the best of our knowledge, this paper is the first to comprehensively explore the impact of various prompt designs on sequential recommendation tasks. Our research extends across six LLMs and three distinct datasets, marking a significant advancement in understanding how prompt strategies influence LLM performance in this field.
    \item We introduce the Dynamic Reflection with Divergent Thinking (\modelname) in a retriever-reranker framework, a novel approach that effectively integrates collaborative signals and the temporal evolution of user preferences into sequential recommendation tasks using LLMs. This framework is specifically designed to be robust against the noise commonly found in datasets, thereby enhancing the reliability of recommendations.
    \item The implementation of our \modelname principle has led to notable improvements in the performance of six different LLMs across three datasets. These significant results underscore the efficacy of our design and highlight the importance of further research in this direction to refine and optimize prompt strategies for sequential recommendation systems.
\end{itemize}
\section{Related Work}
\subsection{LLM for few-shot sequential recommendation}
The work of \cite{wang2023enhancing} utilizes the LLM to generate the reasoning graph along the user attribute and interacted items, and use LLM to guess the next possible item to enrich the original user-item interaction graph. Then, they utilize the conventional GNN to learn the user/item embedding on such an enriched graph for recommendation. For the prompt design, they feed the user history interactions and attributes together to guide the LLM to generate possible reasoning chains among such factors. They also prompt masked user histories, and let the LLM fill in the mask with possible items. They compute the plausible scores measuring the similarity score between the filled-in items and the original items. RecMind~\cite{wang2023recmind} introduces the autonomous agents for multiple recommendation tasks by integrating SQL and database to incorporate item-related information for decision-making. The prompt they adopt for sequential recommendation is the combination of historical interactions and candidate items. They also introduce the self-inspiring reasoning that enables the LLM to reason through multiple paths. \cite{liu2023chatgpt} conduct preliminary case studies on GPT for multiple recommendation tasks. For sequential recommendation, they conduct the zero-shot prompt that is similar to \cite{wang2023recmind}, a few-shot prompt that uses the previous interactions with the second last one as the example, and update the records for the final recommendation.  \cite{liu2023llmrec} also conduct inference on LLM resembling the zero-shot prompt construction. \cite{wang2023zero} introduces a 3-step prompting strategy. They first use LLM to analyze the preference according to the user historical interactions (Step-1), then, they let LLM to pick movies according to the preference analysis (Step-2), and finally, they ask the LLM to rank movies given in the candidate sets (Step-3). The candidate sets are collected according to the cosine-smiliaries between item k-hot encoders constructed from the interacted users. \cite{sanner2023large} utilize completion, zero-shot, and few-shot prompt methods for LLM inference. \cite{hou2023largellmrank} also adopt the zero-shot and few-shot prompting methods. The candidate sets are constructed via random selection or embedding-based model similarity prediction. 

Another direction is to fine-tune the existing LLM for recommendation task. \cite{gao2023chat} also utilize the zero-shot prompt including the user profiles, movie rating, and categories for LLM to select top-k items from candidates. \cite{harte2023leveraging} finetune an LLM by completing the input corpus. The corpus the formalized in a zero-shot prompt manner, that forces the LLM to fill in the last item based on the historical interactions. TALLRec~\cite{bao2023tallrec} constructs the recommendation task as the CTR task, they construct the task corpus using historical interaction as well as the user's ratings, the task is to predict whether the user likes the last item. PALR~\cite{chen2023palr} also fine-tunes LLM using historical interactions. The input corpus are generated in a two-step manner. They first let the LLM summarize the user preferences according to the historical interactions, then, they let the LLM select the next item from candidates. BIGRec~\cite{bao2023bi} first fine-tunes the LLM using the training corpus that are generated in zero-shot prompt format, i.e., input is the historical interactions, and the output is the last item. Then, they compare the L2 distance between the LLM output and item representations for a top-k recommendation. They also incorporate the popularity factor to rescale the L2 distance, such that the more popular item will obtain a smaller distance and, thus, a higher ranking. ReLLa~\cite{lin2023rella} further improves TALLRec by introducing the semantic-related sequences, which are the top-1 closed sequences measured in the space of LLM. They also adopt a zero-shot prompt similar method to construct the prompt. 
\subsection{COT Reasoning in Large Language Models}
Chain-of-Thought is one principle for prompt design, that facilitates the LLM to generate the answer along with the provided reasoning path~\cite{wei2022chain}. \cite{kojima2022large} introduce the Zero-Shot-COT with a naive initialization adding "please think step-by-step" to the prompt. It is a very coarse level design and may lead significant error rate even with the most powerful LLM e.g., ChatGPT. Especially for the recommendation task, where the hidden reasoning varies across different users and different items. \cite{zhang2022automatic} introduce the Auto-COT inspired by~\cite{wei2022chain}, they first cluster the questions into k-clusters, and sample typical questions in each cluster to generate the reasoning chains using the Zero-Shot-COT. Then, they sample such demonstrations into the prompt template in the Manual-COT manner. COT-SC~\cite{wang2022self} is proposed to sample multiple COT reasoning paths and choose the major voted answers as the final answer. \cite{wang2022rationale} proposes the prompt order ensemble and input-rationale ensemble COT methods to improve the LLM inference performance from the ensemble perspective. \cite{li2023making} further improves the COT by introducing the diversified prompt for each question and training an additional path voter to assign the probability of the rationality of each generated reasoning path. They also introduce the strategy to assign the pseudo label of each step, which is included during the optimization procedure of path voters. COT-RR~\cite{he2022rethinking} first sample diverse set of reasoning paths for a given question. Given such reasoning paths and corresponding predictions, they design a post-processing mechanism that checks the faithfulness of reasoning paths according to the retrieved fact from the common knowledge base. IRCOT~\cite{trivedi2022interleaving} interleaves the retrieval and COT in an alternative manner. They first extend the reasoning step of COT, and use the intermediate reasoning step as the query to retrieve the intermediate knowledge. 
\subsection{Self-Verification in Large Language Models}
COT is vulnerable to intermediate mistakes and results in error accumulations.
Instead of fine-tuning an additional LLM as the verifier, recent works tend to design prompts to allow LLMs to self-verify. \cite{ling2023deductive} introduce the natural program to generate the in-context verification step-by-step. \cite{madaan2023self} utilize the very LLM serving feedback for their output and refine their output iteratively. \cite{shinn2023reflexion} adopt one LLM as the evaluator, and another LLM as the feedback giver, and both of them are stored in the memory module. \cite{weng2022large} use the LLM to generate multiple candidate answers, then they rewrite such answers or mask the conditions, which is fed into another LLM for inference with COT. They rank the second LLM's output with the ground truth or masked conditions to give the corresponding ranking score of the original candidates. However, the above methods rely on feedback generated by LLM, which still exists hallucinations within the feedback. \cite{huang2023large} evaluate the intrinsic self-evaluation methods, and point out that such paradigms are not always beneficial to the task. On the contrary, utilizing external tools or human feedback instead of LLM provides strict supervision of the LLM generation. However, in the recommendation task, the sequence is already the human feedback, we can directly utilize such human feedback to correct the LLM generation process. \cite{valmeekam2023can} also evaluate the effectiveness of the self-critic strategy using LLM as the verifier. They conducted empirical studies on the plan generation task, and the results indicate that the self-criticism from LLM degrades the performance compared to that using external feedback as the critic source. 

In summary, both of these two directions do not dig deep into the prompt design for the sequential recommendation. They simply concate the history and the candidates together in a prompt. Current methods have pointed out that the prompt format can vary the performance in a substantial range. In this paper, we focus on the discussion of the effect of prompt design for recommendation performance. And we propose the prompt design principle for the sequential recommendation that dynamically reflects the belief of user preference that deep dive into the user feedback.

\section{Problem Definition}
Given the sequence of item title $s = \{i_1, i_2, \dots, i_s\}$ of user $u$, and a pretrained large language model $LLM$, we design a prompt method $p(s)$ that maximize the ranking performance of $LLM$ over a pre-selected candidate item set $\{c_1, c_2, \dots, c_n\}$. 
\section{Methodology}
\begin{figure}
    \centering
    \includegraphics[width=\textwidth]{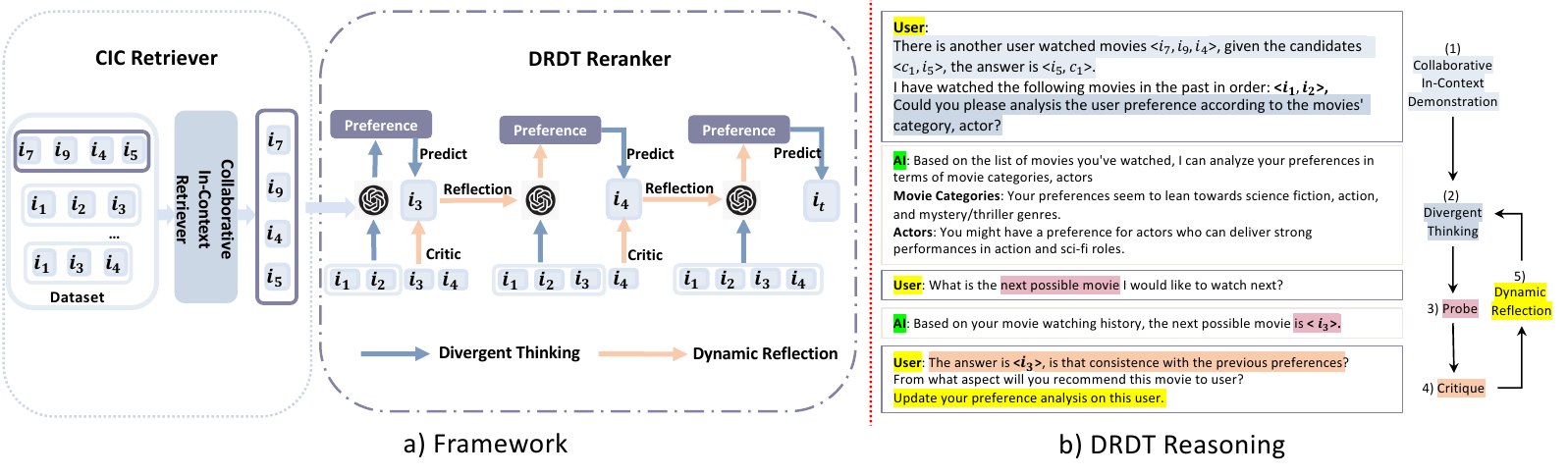}
    \caption{Illustration of \modelname within retriever-reranker framework in a) and the reasoning process of \modelname in b).} 
    \label{fig:framework}
\end{figure}
In this section, we first describe the retriever-reranker framework, specifically designed for recommender systems. Then, we discuss the details of each component, including the collaborative in-context demonstration retriever, the divergent thinking, and the dynamic reflection of user preferences with their feedback.
\subsection{Retriever-Reranker Framework}
One challenge in leveraging LLMs for recommendation tasks is the inability to process the entire dataset during inference. Firstly, excessive noise is a concern. For instance, when recommending movies to a user $i$ who primarily consumes fantasy films, it's crucial to focus on movies relevant to their viewing history. Irrelevant data in the dataset can act as noise, potentially misleading the LLM and impairing its performance. Secondly, the token length limitation of LLMs necessitates the selection of the most informative examples for effective processing. Recently, the use of retrievers in retrieval-augmented generation has gained increasing attention. These retrievers typically leverage semantic distances within the embedding space to identify sentences most similar to a given context as in-context examples. However, in the realm of recommender systems, collaborative information, which focuses on extracting similar purchase behavior patterns from the dataset, is vital. This means that there may be item co-occurrence patterns and noise in two sequences, where the noise results in the dissimilarity between these two sequences, a factor not typically accounted for in standard retrieval models. These considerations highlight the need for a retriever specifically designed for recommendation tasks, one that can effectively identify and utilize relevant collaborative information.

Another significant challenge arises from the mismatch between the typical generation tasks for which LLMs are optimized and the specific needs of recommender systems. In a recommender system, the objective is to have the system identify the top-k items that align with a user's potential interests. This requires the LLM's output to reference existing items rather than generating new pseudo items. Furthermore, users often exhibit a broad spectrum of interests that evolve over time. However, LLMs, being trained primarily for generating the next token based on input queries, lack the explicit capability to analyze and track the temporal evolution of user preferences. This limitation underscores the necessity for a reranker capable of dynamically analyzing user preferences in a temporal manner. Additionally, the vast item set typical in recommender systems makes it impractical to include all items for reranking by an LLM. Many effective methods exist for pre-selecting a subset of candidate items for further reranking. Therefore, our focus shifts to enhancing the reranking capabilities of LLMs over these pre-selected sets of candidate items, ensuring a more targeted and efficient recommendation process.

Formally, our framework comprises a retriever component and a reranker as shown in Figure~\ref{fig:framework}. The retriever's primary function is to identify the most relevant purchase behaviors within the dataset, which serve as in-context references for the subsequent reranking process. The reranker, on the other hand, focuses on two key tasks. Firstly, it extracts collaborative information from the in-context examples provided by the retriever. Secondly, it analyzes user preferences dynamically based on their historical records. The reranker ensures that the recommendations are both relevant to the user's past behavior and reflective of emerging trends or changes in their preferences.

\subsection{Collaborative In-Context Demonstration Retriever}

In-context learning (ICL) has demonstrated effectiveness in improving the inference capabilities of LLMs, as it offers a thinking strategy and a template for answering queries. However, the application of ICL in recommendation tasks presents unique challenges compared to traditional NLP tasks like question answering (QA). In QA, the temporal pattern of the content is not typically critical, whereas, in sequential recommendations, it's essential to consider both the global temporal patterns and the content of the items to make informed decisions. Current implementations of ICL in recommender systems tend to focus solely on extracting purchase behaviors from a given sequence. For example, given a historical sequence  $s=\{i_1,i_2, i_3, i_4\}$, the in-context example might be framed as: \textit{the historical records are } $\{i_1, i_2\}$ \textit{and user purchased} $i_3$. This approach highlights the sequential pattern within the last two interactions, but it does not provide the LLM with sufficient context or collaborative information to effectively recommend the next likely engagement.

To effectively utilize information for the LLM reranker's reference procedure, it is crucial to discern and choose the most relevant data while avoiding noise or distractions. Intuitively, incorporating collaborative information into the LLM helps it understand the likely actions of other users in similar circumstances. We achieve this by using the last item in the user's historical sequence, as an anchor to identify other users who have also interacted with this item as shown in Figure~\ref{fig:framework}. Formally, given the user history $s=\{i_1, i_2, \dots, i_s\}$, we random sample historical records that also end at $\{i_1', i_2' \dots, i_s\}$, as well as their next purchase $i_t$ to construct the in-context example as \textit{There is another user that has history:} $\{i_1', i_2' \dots, i_s\}$, \textit{given the candidate set} $\mathcal{C} = \{c_1, i_t, \dots, c_n\}$, \textit{the possible answer is} $\{i_t, c_i, \dots, c_j\}$, where the possible answer is the pseudo answer that gives the next possible engagement $i_t$ the higher rank. This approach not only provides the LLM with insights about potential next engagements based on collaborative user behavior but also indicates which candidates should be ranked higher. 

\subsection{Divergent Thinking (DT)}

Chain-of-Thought (COT) reasoning has gained prominence in enhancing LLM inference for complex tasks where a direct, one-step answer generation is insufficient. The recommendation task, particularly concerning predicting the next user engagement, presents a similarly intricate scenario that demands a nuanced application of COT. In this context, there are several key challenges to consider in order to effectively guide the LLM in understanding the behavior of users: 1) Multi-aspect Interests: User records often encode a variety of interest aspects. A user's interaction with different items can represent diverse facets of their preferences, ranging from genre preferences in media consumption to specific features in product choices. 2) Ubiquity of Noise Interactions: Not all interactions in a user's history are equally indicative of their true preferences. Some may be outliers or noise, such as accidental clicks or one-time exploratory actions, which can skew the understanding of their interests. Given these challenges, the conventional application of COT, which might involve providing a specific reasoning example or simply appending a "Let's think step by step" prompt, falls short for recommendation tasks.

To tackle the identified challenges in recommendation tasks, we introduce a \textit{Divergent Thinking} approach as shown in Figure~\ref{fig:framework} blue arrows. This method involves initially extracting a high-level abstraction of user preferences across multiple aspects, and then utilizing this abstraction as a guidance for predicting the next engagements. The process begins by querying the LLM (e.g., GPT) to identify the critical aspects users might consider when engaging with a specific item. For instance, in the context of movie watching, aspects such as actors, directors, or genres could be pivotal in a user's decision-making process. By identifying these aspects, we can analyze user preferences from a multifaceted perspective, effectively sidestepping the noise presented in raw input sequences. Formally, for each dataset, we prompt the LLM to generate a list of aspects that users may consider: $\{a_1, a_2, \dots, a_k\}$. We then derive a list of user preference analyses based on these aspects respectively: $\{p_1, p_2, \dots, p_k\}$. These high-level preference analyses are subsequently incorporated into the reranking process during LLM inference.

\subsection{Dynamic Reflection with user feedback (DR) }

Addressing the temporal evolution of user interests is a crucial yet challenging aspect of sequential recommendation, especially when using LLMs. This task is more complex than standard NLP tasks like QA or sentiment analysis. In sequential recommendation, understanding user preferences goes beyond merely arranging behaviors in temporal order due to several intricacies: 1), Temporal Evolution of Interests: User interests are not static; they evolve over time. The factors driving these changes are often unobserved or hidden within the data. A failure to capture this temporal evolution leads to reducing the recommendation process to simplistic methods like majority counting, which do not truly reflect individual user preferences. 2), Alternating Short- and Long-Term Correlations: User behavior often exhibits a mix of short- and long-term interests. For instance, a user might consistently enjoy fantasy movies (long-term interest) but occasionally watch a comedy for a family event (short-term deviation). This interplay between consistent long-term interests and intermittent short-term preferences adds complexity to understanding user preferences over time.

The Divergent Thinking (DT) approach addresses these complexities to a certain degree in understanding user preferences, yet it faces several limitations: First, Difficulty in Preference Analysis: Generating a comprehensive user preference analysis from raw sequences is challenging, especially with complex, evolving sequences. In such scenarios, the LLM may default to simply counting the majority of occurrences, which can oversimplify and misrepresent the true, nuanced preferences of the user. Second, Vulnerability to Hallucination: DT relies on the internal reasoning capabilities of the LLM. If this reasoning is flawed or invalid, it can lead to hallucination – where the LLM generates outputs based on incorrect or unfounded assumptions. As a result, the reranking based on these outputs may lead to inaccurate or misleading recommendations. Third, Decision-making Complexity with Multiple Aspects: While DT provides insights into user preferences from various aspects, it can be difficult for the LLM to prioritize these aspects effectively. Determining which aspects are most influential in predicting the user's next engagement is a complex task that DT does not inherently resolve.

To overcome these limitations, we propose a novel \textit{dynamic reflection (DR)} scheme as shown in Figure~\ref{fig:framework} orange arrows. This method enables the learning of user preferences as they evolve over time, while omitting the accumulations of errors caused by hallucinations. DR mimics the human learning process and involves probing, critiquing, and reflecting. First, \textbf{Inital Subsequence Analysis with DT}: We begin by scrolling back $T$ steps in the user's item sequence, where $T$ is a predefined hyperparameter. For the subsequence $\{i_1, i_2, \dots, i_{s-T}\}$, we apply the DT approach to derive a multi-aspect user preference analysis $\{p_1, p_2, \dots, p_k\}$. Second, \textbf{Engagement Prediction and Reflection}: The LLM is then tasked with predicting the next possible engagement between the actual target item and a randomly chosen candidate item $\{i_{s-T+1}, r\}$. We use $i_{s-T+1}$ as the ground truth for reflection, informing the LLM of the user's actual choice and querying whether its prediction aligns with this reality. The LLM is prompted to reflect on its multi-aspect belief and adjust accordingly. Third, \textbf{Aspect-Specific Analysis Reinforcement}: To enhance the focus on specific aspects of user behavior, we ask the LLM to identify which aspect might have influenced the user's choice of $i_{s-T+1}$ and give more weight to this factor in its analysis. Fourth, \textbf{Stepwise Aspect Importance Update for Reranking}: This probe-critique-reflect procedure is repeated for $T$ steps. With each step, the LLM updates its understanding of the user's behavior pattern, correcting any earlier misjudgments and preventing the accumulation of errors that might arise from hallucinations. After each iteration, we move one step forward, until we reach the last item, and the reranker starts the reranking process for the target item.

\section{Experiments}

\paragraph{Dataset.} In this paper, we adopt the Movielens-1M dataset and the Games and Luxury Beauty categories from the Amazon Review dataset, following common practice in sequential recommendation scenarios. We chose these datasets for several reasons: First, information about movies, games, and beauty products is ubiquitous online, and we hope the existing pre-trained LLMs can provide useful contextual information for such items to help improve recommendation performance. Second, the Amazon dataset is known for its sparsity and cold-start issues, which necessitate research on zero/few-shot recommendations leveraging supportive information from LLMs. We report the statistics of the datasets in Table~\ref{tab:data_stat}.

\begin{table}[]
    \caption{Dataset Statistics}
    \small
    \centering
    \begin{adjustbox}{width=\textwidth}

    \begin{threeparttable}
        \begin{tabular}{c|c|c|c|c|c|c}
        \toprule
              Dataset &\# Users & \# Items & \# Interactions & Avg. Actions of Users &  Avg. Actions of Items & Sparsity\\
             \midrule
             ML-1M & 6,041 & 3,707 & 1,000,209 & 165.60 & 269.89 & 95.53\%\\
             Games & 50,567 & 16,860 & 389,718 & 7.71 & 23.12 & 99.95\%\\
             Luxury & 2,028 & 936 & 18,005 & 8.88 & 19.26 &  99.05\% \\
             \bottomrule
         \end{tabular}
    \end{threeparttable}
    \end{adjustbox}
    \label{tab:data_stat}
\end{table}

\paragraph{Metrics.}
Following the common practice in sequential recommendation~\cite{SASRec, wang2022contrastvae, Bert4Rec, DuoRec}, we adopt the Recall@K and NDCG@K as the metrics of ranking performance of recommender systems, where K=10, 20. 
\paragraph{Baseline Methods.}

The zero-shot/few-shot sequential recommendation using pre-trained LLMs is still under-explored. Here, we compare our methods with common strategies adopted in ~\cite{liu2023llmrec, hou2023largellmrank}, including Plain-text, ICL, and COT. Specifically, Plain-text uses historical interactions as sentences to construct the prompt, allowing the LLM to rank candidate items, which include the target items. ICL uses the historical interactions before the second-to-last item as the query sentence, and the second-to-last item as an answer to construct the in-context example. The query is then updated to include historical interactions, prompting the LLM to rank candidate items, including the target items. COT is the zero-shot COT that adds "Let's think step-by-step" at the end of the prompt to guide the LLM in ranking along the intermediate reasoning paths.
\paragraph{Implementation Details.}

We use FastChat to launch pre-trained language models, including Vicuna-7b-16k, Vicuna-13b-16k, Openchat-7b, Longchat-7b-16k, Longchat-13b-16k, and Yarn-Mistral-7b-128k, which serve as the reranker module in our framework. We also adopt the API-based GPT-Turbo-3.5 as a Reranker for comparison. We randomly sample 200 users from each dataset and use the target items along with 19 randomly sampled items to construct the candidate item set. It is worth noting that we randomly pick the position of the target item in the candidate set to prevent the LLM from recognizing a shortcut to the answer.

\subsection{Overall Comparison}

In this section, we present a performance comparison between our reasoning principle and existing methods like plain-text, ICL, and COT, using pre-trained large LLMs including Vicuna-7b, Vicuna-13b, Openchat-7b, Mistral-7b, Longchat-7b, and GPT-Turbo-3.5. The experimental results are detailed in Table~\ref{tab:exp-overall_comp}. Based on these numerical results, we observe the following: 

First, our strategy, which utilizes dynamic reflection and divergent thinking on inference to practice to predict the future, significantly outperforms plain-text, ICL, and COT across all datasets. Our approach successfully enhances the ranking performance of all LLMs on all datasets with respect to almost all metrics. These results underscore the effectiveness of our reasoning strategy. Moreover, our strategy enables a relatively smaller LLM, such as Openchat-7b, to surpass the performance of GPT-Turbo-3.5 with plain-text prompting across the ML-1M, Games, and Luxury datasets in terms of Recall@10, NDCG@10, and NDCG@20 metrics. These findings underscore the significant impact that the choice of reasoning strategy can have on an LLM's performance. With thoughtful design, even a 7-billion parameter LLM can outperform larger models like GPT-Turbo-3.5. This demonstrates the potential for more strategic reasoning design to optimize the performance of LLMs in various tasks, especially in zero/few-shot ranking or when fine-tuning LLMs for recommendation scenarios. We hope our paper will inspire further research in this area, focusing on the development of innovative reasoning strategies tailored to enhance the capabilities of LLMs in recommendation and other complex applications.

Regarding the ICL and COT strategies, we observed that different models show a preference for different prompting strategies across various datasets. For instance, Vicuna-13b using ICL achieved the second-best performance on the ML-1M and Games datasets. However, its performance was worse than the plain-text strategy on the Luxury dataset. A similar pattern is evident with the COT strategy; Vicuna-13b-COT underperformed plain-text on the ML-1M dataset but surpassed it on the Games and Luxury datasets. For Mistral-7b and Longchat-7b, both ICL and COT generally underperformed compared to plain-text, except on the Luxury dataset. These observations suggest that ICL and COT may be more suitable for certain LLMs and specific datasets. Typically, models that already perform reasonably well with plain-text benefit more from ICL. This could be because these LLMs have better-encoded knowledge, and ICL assists in understanding the ranking task and retrieving useful information. For LLMs with better overall performance, such as Openchat and GPT-Turbo-3.5, ICL tends to outperform COT, and COT often performs worse than plain-text. This could be attributed to the highly personalized nature of recommendation tasks, where there may not be a universal reasoning path of COT suitable for all users. Additionally, the input sequences are complex and evolve dynamically, making it challenging for LLMs to analyze user preferences over time without a defined reasoning path. They tend to make recommendations based on similarity comparisons. From another perspective, models with weaker plain-text performance, such as Mistral-7b and Longchat-7b, generally show that both COT and ICL underperform plain-text, except in the case of the Luxury dataset. This trend might stem from these models' challenges in comprehending the task through in-context examples and their difficulty in autonomously generating a coherent reasoning path. These may inadvertently introduce more noise into the process, rather than providing useful insights. In contrast, our reasoning strategy combines the strengths of both ICL and COT, guiding the LLM's reasoning in a temporal manner to dynamically analyze user preferences. Even with Mistral-7b, \modelname managed to provide reasonable performance with a maximum of 209.38\% improvements on the Luxury dataset. This approach consistently improves performance significantly across different models and datasets.

\begin{table*}[t]
    \caption{Overall Comparison. }
    \label{tab:exp-overall_comp}
    \small
    \centering
    \begin{adjustbox}{width=\textwidth}
    \begin{threeparttable}
    {
    \begin{tabular}{l|cccc|cccc|cccc}
         \toprule
         \multirow{2}{*}{\textbf{Models}} 
         & \multicolumn{4}{c|}{\textbf{ML-1M}} & \multicolumn{4}{c|}{\textbf{Games}} & \multicolumn{4}{c}{\textbf{Luxury}}\\
         & R@10 & @ R@20 & N@10 & N@20 & R@10 & @ R@20 & N@10 & N@20 & R@10 & @ R@20 & N@10 & N@20 \\
         
         \midrule
         \textbf{Vicuna-7b} & 0.3700 & 0.6400 & 0.1733 & 0.2421 & 0.3700 & \underline{0.6150} & 0.1783 & 0.2405 & \underline{0.3450} & 0.4800 & 0.1534 & 0.1882\\
        \textbf{Vicuna-7b-ICL} & \underline{0.4450} & \underline{0.7800} & \underline{0.2183} & \underline{0.3024} & 0.4000 & 0.5950 & \underline{0.2102} & \underline{0.2598} & 0.3200 & 0.4500 & \underline{0.1589} & 0.1927 \\
         \textbf{Vicuna-7b-COT} & 0.3700 & 0.6850& 0.1667& 0.2461 & \underline{0.4100} & \underline{0.6150} & 0.1948 & 0.2467 & 0.3400 & \underline{0.5000} & 0.1562 & \underline{0.1979} \\
         \textbf{Vicuna-7b-\modelname} & \textbf{0.5250} & \textbf{0.8450} & \textbf{0.2698} & \textbf{0.3500} & \textbf{0.4300} & \textbf{0.6900} & \textbf{0.2361} & \textbf{0.3023} & \textbf{0.4550} & \textbf{0.6300} & \textbf{0.2418} & \textbf{0.2867}  \\
         \textbf{Improvement (\%)} & 17.98 & 8.33 & 23.59 & 15.74 & 4.88 & 12.20 & 12.32 & 16.36 & 31.88 & 26.00 & 52.17 & 44.87\\
         \midrule
         
         \textbf{Vicuna-13b} & \textbf{0.5550} & 0.8000 & 0.2855 & 0.3475 & \underline{0.4400} & 0.6700 & 0.2269 & 0.2855 & 0.2800 & 0.4600 & 0.1525 & 0.1995 \\
         \textbf{Vicuna-13b-ICL} & \underline{0.5450} & \underline{0.8050} & \underline{0.2885} & \underline{0.3541} & 0.4150 & \underline{0.6950} & 0.2293 & \underline{0.2988} & 0.2550 & 0.4100 & 0.1356 & 0.1755 \\         
         \textbf{Vicuna-13b-COT} & 0.4850 & 0.7600& 0.2354 & 0.3044 & 0.4400 & 0.6600 & \underline{0.2344} & 0.2895 & \underline{0.3100} & \underline{0.5000} & \underline{0.1563} & \underline{0.2057}\\
        \textbf{Vicuna-13b-\modelname} & \underline{0.5450} & \textbf{0.8150} & \textbf{0.3153} & \textbf{0.3824} & \textbf{0.4500} & \textbf{0.7150} & \textbf{0.2731} & \textbf{0.3399} & \textbf{0.4450} & \textbf{0.6600} & \textbf{0.2608} & \textbf{0.3164}\\
        \textbf{Improvement (\%)} & -1.80 & 1.24 & 9.29 & 7.99 & 2.27 & 2.88 & 16.51 & 13.76 & 43.55 & 32.00 & 66.86 & 53.81 \\
         \midrule
         
         \textbf{Openchat-7b} & 0.4200 & 0.6650 & 0.2075 & 0.2684 & 0.1450 & 0.3700 & 0.0702 & 0.1264 & 0.2950 & 0.5300 & 0.1341 & 0.1945\\
         \textbf{Openchat-7b-ICL} & \underline{0.5100} & \underline{0.8300} & \underline{0.2734} & \underline{0.3542} & \underline{0.4200} & \underline{0.6900} & \underline{0.2355} & \underline{0.3038} & \underline{0.4300} & \underline{0.5900} & \underline{0.2229} & \underline{0.2642} \\
         \textbf{Openchat-7b-COT} & 0.4350 & 0.6950 & 0.2209 & 0.2867 & 0.2200 & 0.5050 & 0.1096 & 0.1806 & 0.2300 & 0.4500 & 0.1151 & 0.1705\\         
         \textbf{Openchat-7b-\modelname} & \textbf{0.6350} & \textbf{0.9000} & \textbf{0.3335} & \textbf{0.4016} & \textbf{0.4800} & \textbf{0.7150} & \textbf{0.3069} & \textbf{0.3651} & \textbf{0.4550} & \textbf{0.6500} & \textbf{0.2589} & \textbf{0.3091} \\
         \textbf{Improvement (\%)} & 24.51 & 8.43 & 21.98 & 13.38 & 14.28 & 3.62 & 30.32 & 20.18 & 5.81 & 10.17 & 16.15 & 16.99
          \\
         \midrule
         
         \textbf{Mistral-7b} & \underline{0.2350} & \underline{0.3350} & \underline{0.1109} & \underline{0.1362} & \underline{0.2000} & \underline{0.2000} & \underline{0.1094} & \underline{0.1094} & 0.1000 & 0.1100& 0.0590 & 0.0617 \\
         \textbf{Mistral-7b-ICL} & 0.1150 & 0.2000 & 0.0497 & 0.0711 & 0.1700 & 0.1700 & 0.1072 & 0.1072 & 0.1350 & 0.1500 & \underline{0.0995} & \underline{0.1034}\\
         \textbf{Mistral-7b-COT} & 0.1950 & 0.3050 & 0.0937 & 0.1220 & 0.1750 & 0.1750 & 0.0915 & 0.0915 & \underline{0.1450} & \underline{0.1600} & 0.0991 & 0.1030\\
         \textbf{Mistral-7b-\modelname}& \textbf{0.3750} & \textbf{0.6300} & \textbf{0.1842} & \textbf{0.2489} & \textbf{0.4450} & \textbf{0.5750} & \textbf{0.2462} & \textbf{0.2797} & \textbf{0.4100} & \textbf{0.4950} & \textbf{0.2512} & \textbf{0.2736}\\
         \textbf{Improvement (\%)} & 59.57 & 88.06 & 66.10 & 82.75 & 122.50 & 187.50 & 125.05 & 155.67 & 182.76 & 209.38 & 152.46 & 164.60 \\
         \midrule


        \textbf{Longchat-7b} & \underline{0.3900} & \underline{0.6450} & \underline{0.1808} & \underline{0.2448} & \underline{0.3450} & \underline{0.5700} & \underline{0.1653} & \underline{0.2223} & 0.2350 & 0.3350 & 0.0976 & 0.1230\\
        \textbf{Longchat-7b-ICL} & 0.3250 & 0.5450 & 0.1555 & 0.2114 & 0.3150 & 0.5000 & 0.1523 & 0.1995 & 0.2400 & 0.3000 & 0.1129 & 0.1286\\
        \textbf{longchat-7b-COT} & 0.3750 & 0.5950 & 0.1739 & 0.2294 & 0.2650 & 0.4950 & 0.1277 & 0.1860 & \underline{0.2600} & \underline{0.3750} & \underline{0.1289} & \underline{0.1592}\\
        \textbf{Longchat-7b-\modelname} & \textbf{0.4650} & \textbf{0.8400} &  \textbf{0.2063} & \textbf{0.3011} & \textbf{0.4150} & \textbf{0.6200} & \textbf{0.2050} & \textbf{0.2566} & \textbf{0.3800} & \textbf{0.5300} & \textbf{0.2047} & \textbf{0.2433}\\
        \textbf{Improvement (\%)} &19.23 & 30.23 & 14.10 & 23.00 & 20.29 & 8.77 & 24.02 & 15.43 & 46.15 & 41.33 & 58.81 & 52.83
     \\
        \midrule

        
        \textbf{GPT-Turbo-3.5} & 0.5450 & \textbf{0.9200} & 0.2688 & 0.3631 & 0.4400 & \textbf{0.7450} & 0.2499 & 0.3275 & 0.4350 & \textbf{0.7000} & 0.2260 & 0.2924\\
        \textbf{GPT-Turbo-3.5-ICL} & \underline{0.6600} & \textbf{0.9200} & \underline{0.3829} & \underline{0.4483} & \underline{0.5050} & 0.7100 & \underline{0.3106} & \underline{0.3618} & \underline{0.5150} & \underline{0.6750} & \underline{0.3049} & \underline{0.3464}\\
        \textbf{GPT-Turbo-3.5-COT} & 0.5050 & 0.9000 & 0.2714 & 0.3694& 0.4400 & 0.7250 & 0.2664 & 0.3371 & 0.3800 & 0.6250 & 0.2030 & 0.2653\\
        \textbf{GPT-Turbo-3.5-\modelname} & \textbf{0.7550} & \underline{0.9100} & \textbf{0.4630} & \textbf{0.5031} & \textbf{0.5450} & \underline{0.7300} & \textbf{0.3435} & \textbf{0.3898} & \textbf{0.5600} & 0.6700 & \textbf{0.3190} & \textbf{0.3482} \\
        \textbf{Improvement (\%)} & 14.39 & -1.09 & 20.91 & 12.22 & 7.92 & -2.01 & 10.59 & 7.74 & 8.74 & -4.29 & 4.62 & 0.52 \\

         \bottomrule
    \end{tabular}
    }
    \end{threeparttable}
    \end{adjustbox}
\end{table*}
\subsection{Ablation Study}

\begin{figure}
    \centering
    \includegraphics[width=0.8\textwidth]{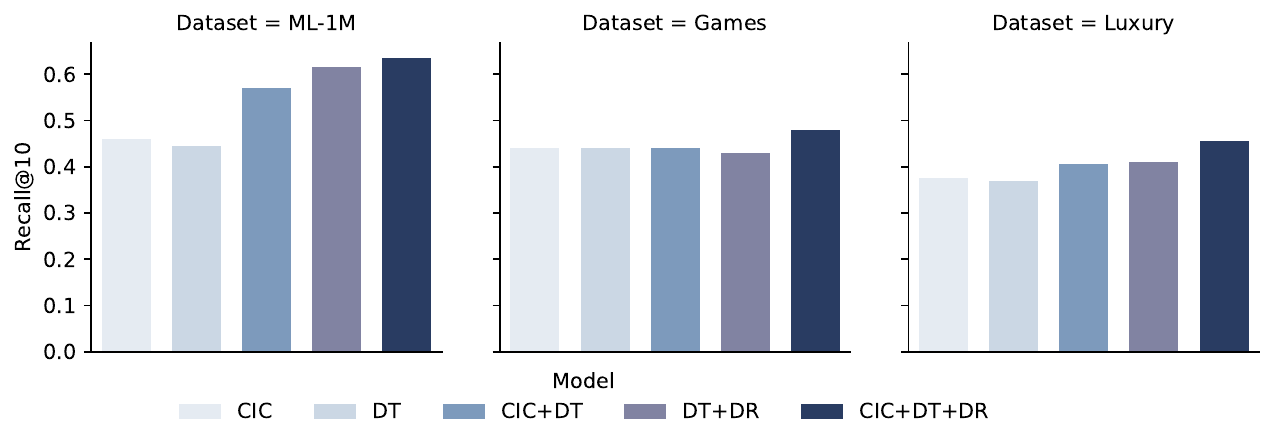}
    \caption{Performance of Openchat-7b with different components on three datasets.} 
    \label{fig:openchat}
\end{figure}

\begin{figure}
    \centering
    \includegraphics[width=0.8\textwidth]{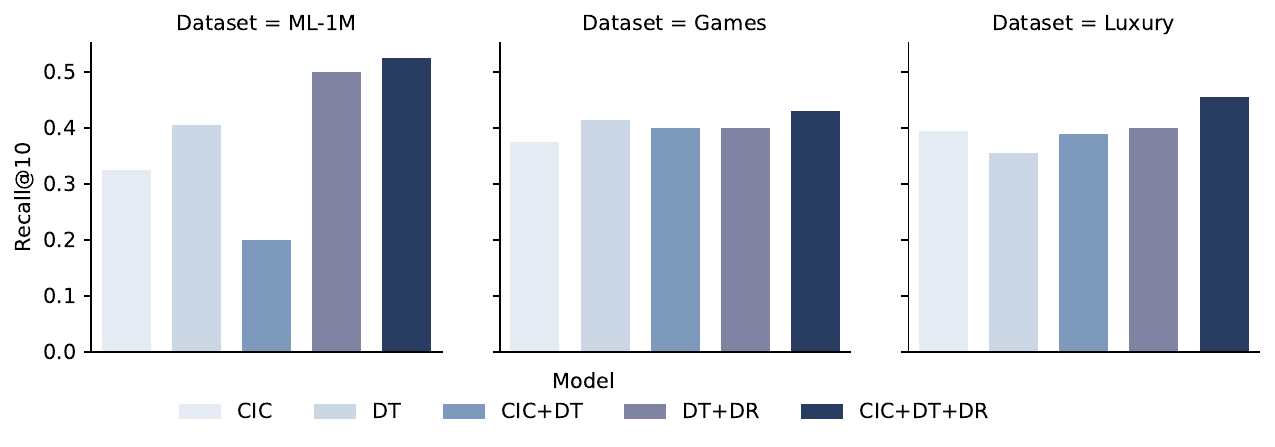}
    \caption{Performance of Vicuna-7b with different components on three datasets.}
    \label{fig:vicuna7b}
\end{figure}

\begin{figure}
    \centering
    \includegraphics[width=0.8\textwidth]{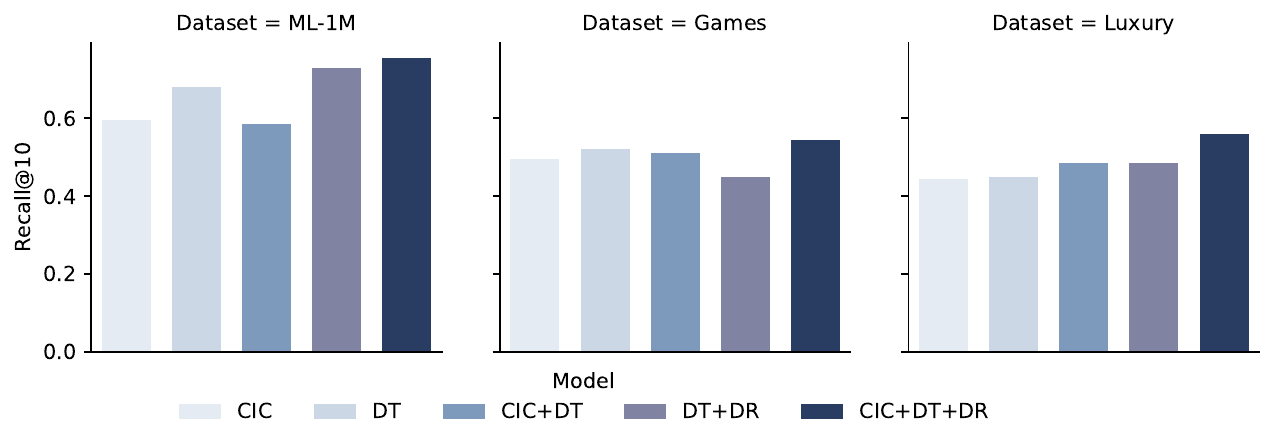}
    \caption{Performance of GPT-Turbo-3.5 with different components on three datasets.}
    \label{fig:gpt-turbo-3.5}
\end{figure}

In this section, we carry out an ablation study to assess the effectiveness of each module within our framework.  Additionally, we report empirical results for Openchat-7b, Vicuna-7b, and GPT-Turbo-3.5 on the ML-1M, Games, and Luxury datasets in Figures~\ref{fig:openchat}, ~\ref{fig:vicuna7b}, and ~\ref{fig:gpt-turbo-3.5} respectively. From these figures, we make several observations:

\paragraph{Importance of CIC, DT, and DR:} Collaborative In-Context (CIC), Divergent Thinking (DT), and Dynamic Reflection (DR) are vital for achieving optimal performance. Removing DR leads to a significant drop in performance, and a similar decline is observed when CIC/DT is excluded. This highlights the crucial role these components play in the overall effectiveness of our model.

\paragraph{Impact of CIC combined with DT on Different Datasets:} While combining DT with CIC consistently improves performance on the Luxury dataset, it has a lesser or even negative impact on the ML-1M and Games datasets. This suggests that CIC does not always aid LLM reasoning; in some cases, it may introduce additional noise or distract the LLM. This observation underscores the importance of Dynamic Reflection in controlling performance analysis through probing. Without such a regulatory mechanism, the combination of CIC and DT can lead to performance deterioration compared to using either CIC or DT alone.

\paragraph{Effectiveness of DT Combined with DR:} The integration of DT with DR consistently enhances performance on the ML-1M and Luxury datasets but has a lesser impact on the Games dataset. This may indicate that collaborative information plays a more significant role compared to the dynamic patterns of user behavior evolution in certain contexts.

\subsection{Hyperparameter Sensitivity}

In this section, we delve into the analysis of Recall@10 and NDCG@10 metrics for Openchat-7b, GPT-Turbo-3.5, and Vicuna-7b on three datasets, focusing on the impact of varying the number of dynamic reflection steps. Here, a single step of dynamic reflection is defined as one cycle of prediction probing and analysis reflection using a ground-truth item at a specific position. The number of steps corresponds to the frequency of probing. For instance, a 3-step dynamic reflection involves using the historical sequence $\{i_1, i_2, \dots, i_{s-4}\}$ for user preference analysis from different aspects, then using item $i_{s-3}$ as the answer for the LLM to reflect on their belief, and so forth.

\begin{figure}
    \centering
    \includegraphics[width=0.8\textwidth]{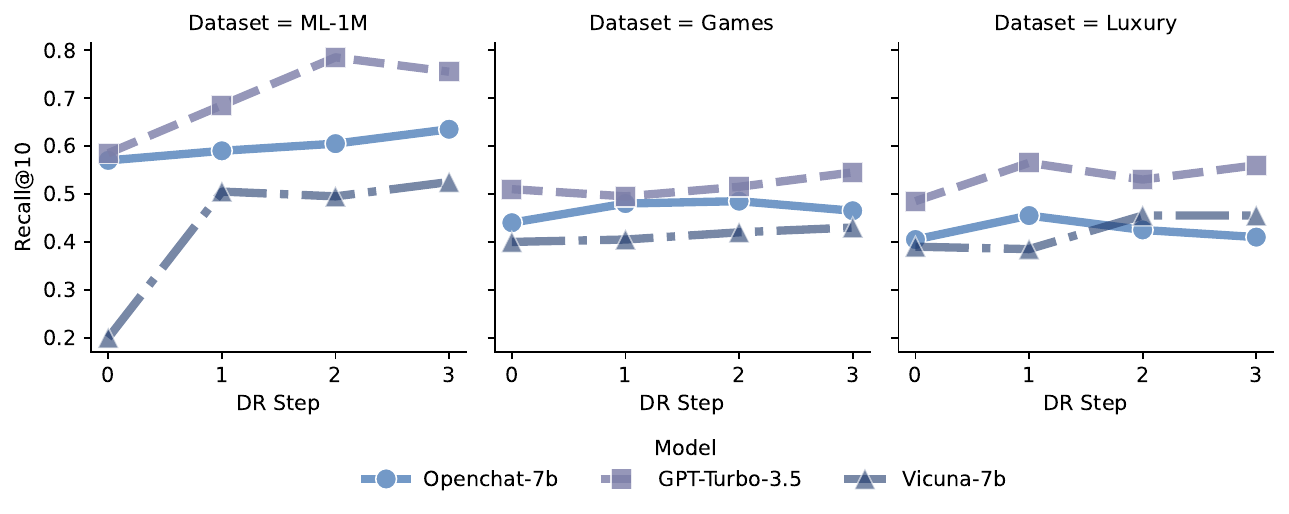}
    \caption{Recall@10 of three LLMs with different DR steps on three datasets.}
    \label{fig:hyper-recall}
\end{figure}

\begin{figure}
    \centering
    \includegraphics[width=0.8\textwidth]{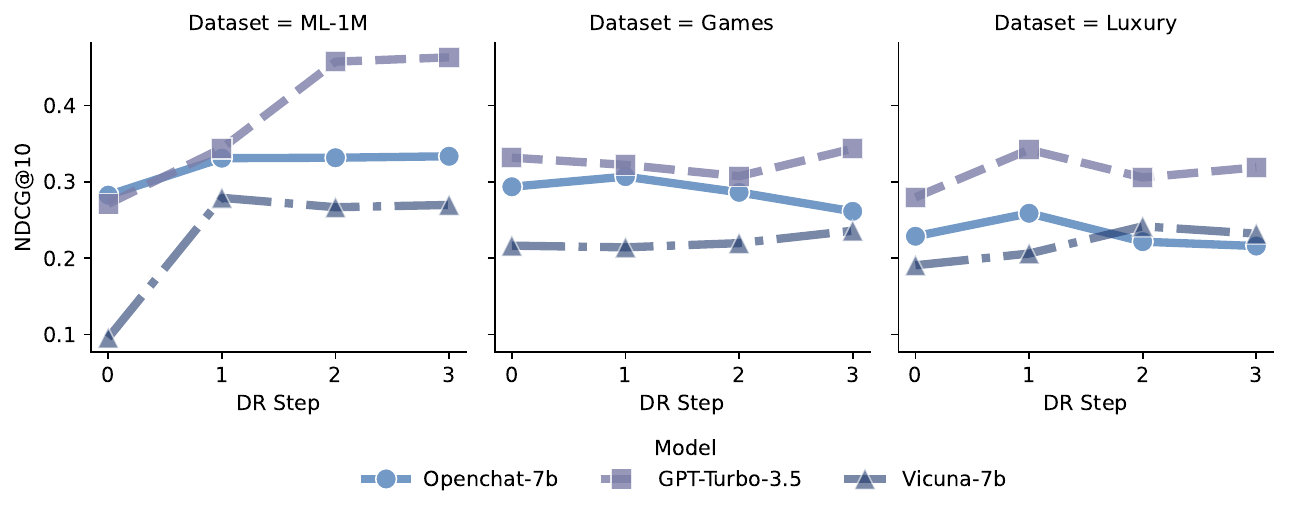}
    \caption{NDCG@10 of three LLMs with different DR steps on three datasets.}
    \label{fig:hyper-ndcg}
\end{figure}

The numerical results are presented in Figures~\ref{fig:hyper-recall} and ~\ref{fig:hyper-ndcg}. Due to constraints such as the context length of the LLMs and the length of user sequences, we have limited our reflection steps to a maximum of 3. Observations from these figures show an increasing trend in performance metrics with the number of reflection steps. This trend underscores the effectiveness of dynamic reflection, where ground-truth items from user behaviors are utilized before making predictions on future engagements. 

\subsection{Case Study}
\begin{figure}
    \centering
    \includegraphics[width=\textwidth]{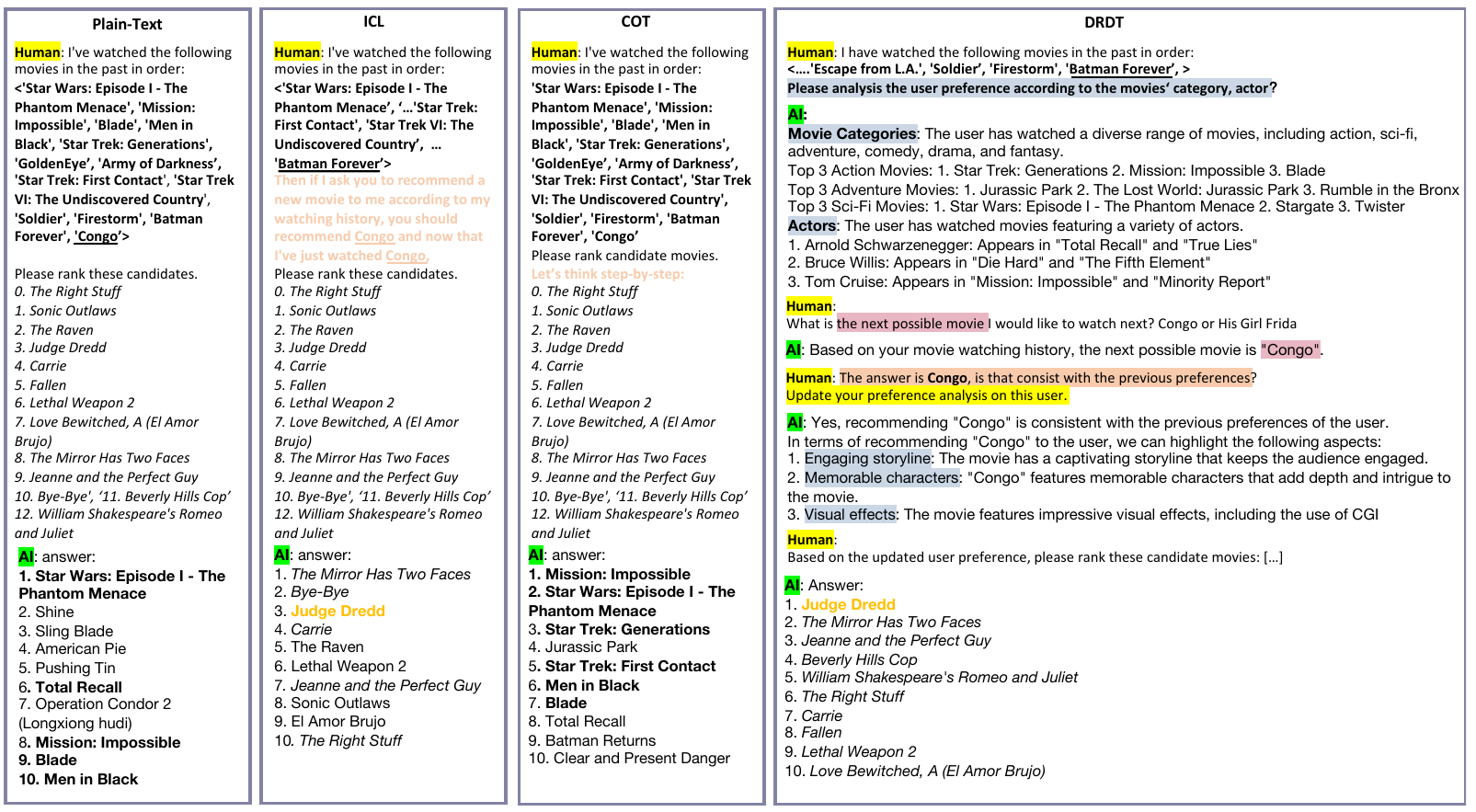}
    \caption{Case Study}
    \label{fig:case}
\end{figure}

In this section, we compare the ranking lists generated by Openchat-7b using \modelname with other reasoning principles. For reference, we consider the plain-text approach, where the historical sequence is used as the query and the LLM is asked to rerank the candidates directly. In the ICL approach, the sequence before the penultimate item is used as the query, guiding the LLM to recommend the second-to-last item to construct the in-context example. The sequence is then updated to include the penultimate item for reranking the candidates. Additionally, we apply zero-shot COT for comparison, given the lack of a universal reasoning path for personalized preferences, and allow the LLM to autonomously determine the reasoning path. We report the results on Figure\ref{fig:case}. Our observations from comparing the outputs are as follows:

1), The superiority of \modelname: The \modelname effectively improved the ranking of the ground-truth item \textit{Judge Dredd} to the first position, surpassing other prompting strategies. This is achieved by posing specific questions to the LLM, enabling it to retrieve more relevant information about the items for recommendation. Additionally, the LLM incorporates more aspects into the user preference analysis through the probe predictions, aiding the final reranking process.
2), Limitations of Plain-Text and COT: Both plain text and COT strategies struggled to accurately understand the task of reranking candidate items. They tend to regenerate historical interactions, as these are the semantically most similar items. COT, in particular, failed to develop a coherent reasoning path, defaulting to retrieving items similar to those in the history, rather than effectively reranking the candidate items.
3), Effectiveness of ICL: ICL helped the LLM understand the task more effectively. The items generated are mostly from the candidate set, and the target items are ranked higher. This indicates that while ICL assists in task comprehension, it may not fully capture the nuances required for optimal reranking in personalized recommendation scenarios.
These insights highlight the strengths and limitations of various prompting strategies and underscore the effectiveness of \modelname in enhancing the LLM's performance in sequential recommendation tasks.

\section{Conclusion}

In this paper, we explore the significant effects of different prompting strategies on the reasoning capacity of Large Language Models (LLMs) for sequential recommendation tasks. We propose a Retriever-Reranker framework and instantiate the retriever with a collaborative in-context demonstration retriever that collects collaborative sequences from the dataset to facilitate recommendations. We introduce divergent thinking, which abstracts high-level user preferences from multiple aspects, and dynamic reflection that mimics the human learning procedure, including a probing-critiquing-reflecting iteration on user preference analysis to encourage the LLM to think in a temporal manner. This reasoning principle successfully improves the LLM's ranking performance substantially without the need for fine-tuning the LLM. We hope this paper could inspire the investigation of maximizing the reasoning capacity of existing LLM for sequential recommendation scenarios. 
\bibliographystyle{abbrvnat}
\bibliography{reference}
\end{document}